\documentclass{article}
\usepackage{spconf,amsmath,epsfig,amsfonts,amssymb}

\newcommand{\Real}{\mathbb{R}}

\def \bS{\mathbf{S}}

\def \bP{\mathbf{P}}

\def \bX{\mathbf{X}}
\def \bY{\mathbf{Y}}

\def \bphi{\mathbf{\Phi}}

\def \bGamma {\mathbf{\Gamma}}

\def \bN{\mathbf{N}}
\def \bY{\mathbf{Y}}

\def \b1 {\mathbf{1}}

\title{Robust Hydraulic Fracture Monitoring (HFM) of Multiple Time Overlapping Events Using a Generalized Discrete Radon Transform}
\name{Gregory Ely and Shuchin Aeron }
\address{{Tufts University, Dept. of ECE, Medford MA}}
\begin{document}
\ninept
\newtheorem{condition}{\indent \bf Condition}[section]
\newtheorem{property}{\indent \bf Property}[section]
\newtheorem{defn}{\indent \bf Definition}[section]
\newtheorem{conj}{\indent \bf Conjecture}[section]
\newtheorem{cor}{\indent \bf Corollary}[section]
\newtheorem{lem}{\indent \bf Lemma}[section]
\newtheorem{claim}{\indent \bf Claim}[section]
\newtheorem{thm}{\indent \bf Theorem}[section]
\newtheorem{prop}{\indent \bf Proposition}[section]
\newtheorem{remark}{\indent \bf Remark}[section]
\newtheorem{example}{\indent \bf Example}[section]

\maketitle

\vspace{-1mm}
\begin{abstract}
\vspace{0 mm}

In this work we propose a novel algorithm for multiple-event localization for Hydraulic Fracture Monitoring (HFM) through the exploitation of the sparsity of the observed seismic signal when represented in a basis consisting of space time propagators. We provide explicit construction of these propagators using a forward model for wave propagation which depends non-linearly on the problem parameters - the unknown source location and mechanism of fracture, time and extent of event, and the locations of the receivers. Under fairly general assumptions and an appropriate discretization of these parameters we first build an over-complete dictionary of generalized Radon propagators and assume that the data is well represented as a linear superposition of these propagators. Exploiting this structure we propose sparsity penalized algorithms and workflow for super-resolution extraction of time overlapping multiple seismic events from single well data.
\end{abstract}

\vspace{-2mm}
\section{Introduction}
\label{sec:intro}
\vspace{-2mm}

Accurate seismic hydraulic fracturing monitoring (HFM) can mitigate many of the environmental impacts by providing a clear real-time image of where the fractures are occurring outside of the shale and how efficiently they are formed within the gas deposit. Although simple in principle, real time monitoring of hydraulic fracturing is extremely difficult to perform successfully due to high noise levels generated by the pumping equipment, anisotropic propagation of seismic waves through shale, and the multi-layered stratigraphy leading to complex seismic ray propagation, \cite{Eisner_TLE09,Aki_Book,Shearer_Book}. In addition the complexity of the source mechanism affects the wave polarization at the 3-axis seismometers, \cite{Chapman_GJI012} introducing extra parameters in the system.

\vspace{-2mm}
\section{Problem Set-up}
\label{sec:setup}
\begin{figure}[ht]
\centering \makebox[0in]{
    \begin{tabular}{c}
      \includegraphics[width = .4\textwidth]{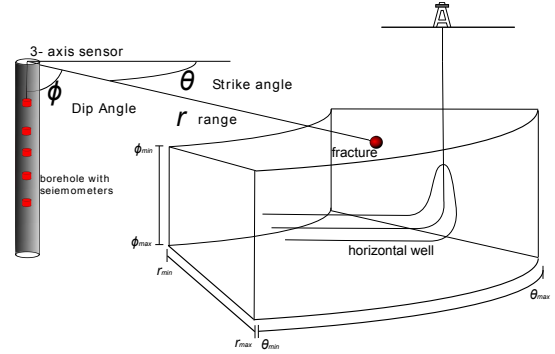}
      \end{tabular}}
  \caption{This figure shows the typical hydraulic fracturing setup in which a seismic array detects the emitted signal from a fracture occurring within the search volume.}
  \label{fig:setup}
\end{figure}

\begin{figure*}[ht]
\centering \makebox[0in]{
    \begin{tabular}{c}
\includegraphics[width=.78\textwidth, height = 1.65 in]{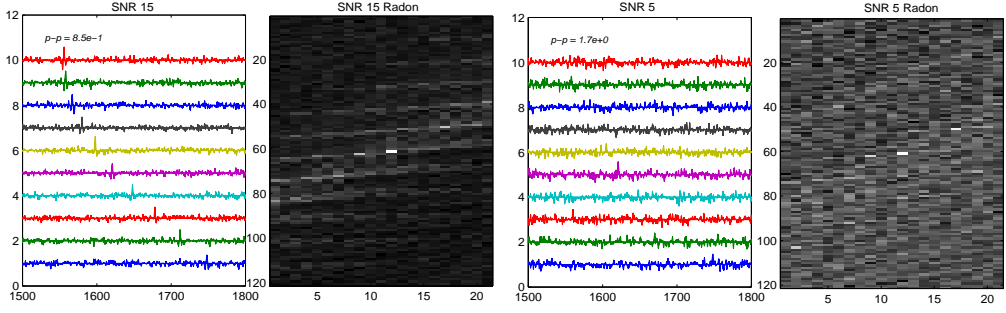}
\end{tabular}}
\caption{This figure shows the traces for a synthetic $\rho$-wave across for two SNRs: 15 and 5. Also shown are the (generalized) Radon transforms. Note the signal sparsity in the Radon domain which is concentrated at a single position index and among a few time indices.}
\label{fig:GDRT_example}
\label{fig:snrTraces}
\end{figure*}

\vspace{-2mm}
The physical set-up is shown in Figure \ref{fig:setup}. A typical seismic array of (say) $L$ three-axis seismometers sample the instantaneous displacement across the three axes at a sampling rate between 1-8 kHz.  The $j^{th}$ detector in the array records a trace, see Figure \ref{fig:snrTraces}, along the three-axes that will be a combination of $m$ seismic events observed in noise which can be effectively modeled as Additive White Gaussian Noise (AWGN), see \cite{Liu_IGARSS09}. Therefore we have,
\begin{align}
\label{eq:Yj}
\bY_j(t) =  \sum_{i=1}^{m}((\bS_{ij}(t) + \bN(t))
\end{align}
where,
\begin{align}
\bY_j(t)=
\begin{bmatrix}
  Y_{jx}(t) \\
  Y_{jy}(t)\\
  Y_{jz}(t) \\
\end{bmatrix}
\;
\bS_{ij}(t)=
\begin{bmatrix}
  S_{ijx}(t) \\
  S_{ijy}(t)\\
  S_{ijz}(t) \\
\end{bmatrix}
\end{align}
In the most general case of anisotropic formation $\bS_{ij}(t)$ is a combination of compressional ($\rho$) wave component and shear component which further consists of vertical shear component, $s_v$, and horizontal shear component, $s_h$, as incident on the detector along the three axes, i.e.,
\begin{align}
\label{eq:Sij}
\bS_{ij}(t) = \bS_{ij\rho}(t) +\bS_{ijs_h}(t) +\bS_{ijs_v}(t)
\end{align}
For a given a wave type such as $\rho$, the signal at the seismometer induced by the wave, $\bS_{ij\rho}(t)$, is completely determined by the signal waveform of the compressional wave at the source, $s_{i\rho}(t)$, the arrival time for the compressional wave of the $i^{th}$ event at the $j^{th}$ detector, $\tau_{\rho_{ij}}$, the compressional polarization unit vector, $\bP_{ij_\rho}$, representing the direction of particle movement at the detector. Usually the signal waveform $s_{i\rho}(t)$ is time compact due to finite duration of the seismic event. Therefore the signal at the seismometer is the signal at the source, $s_{i\rho}(t)$, delayed by a time, $\tau_{\rho_{ij}}$, and projected onto the three axes by the polarization unit vector, $\bP_{ij_\rho}$. Therefore, mathematically we can write,
 \begin{align}
\label{eq:SijP}
 \bS_{ij_\rho}(t) = \delta(t-\tau_{\rho_{ij}})\ast (\bP_{ij_\rho} s_{ij_\rho}(t)) = \bP_{ij_\rho} s_{ij_\rho}(t-\tau_{\rho_{ij}})
\end{align}
The time and location of the event, say $\{t_i,l_i\}$, and velocity profile of the stratigraphy and the seismic source mechanism completely determine the arrival time of the wave and the polarization vector at each of the detector. This is expressed in terms of a forward model function,
 \begin{align}
 \label{eq:bPij}
 \{ \bP_{ij_\rho},\tau_{\rho_{ij}}\} = f_j(t_i,\underset{l_i}{\underbrace{{\theta_i,\phi_i,r_i)}}})
 \end{align}
where we have characterized the location of the event at $l_i$ in terms of the strike, dip and range relative to an absolute co-ordinate system.  Given that the seismic event at a one location will not affect the signal emitted from a seismic event at another location, we can assume that the observed data $\bY_j(t)$ at the $j-$th receiver resulting from events that emit all three wave types can be viewed as a linear combination of signals observed in AWGN.  In this work we will focus on using the compressional waves for HFM as these waves typically have higher amplitude than the shear waves. However, our approach can be easily extended to handle the processing of shear waves. Using  equation~(\ref{eq:bPij}), we can rewrite the Equation~(\ref{eq:Yj}) for the observed trace due to compressional waves, $\bY_j(t)$ as follows,
\begin{align}
\bY_j(t)=\sum^{m}_{i=1} \bP_{ij_\rho} s_{i_\rho}(t-\tau_{\rho_{ij}}) + \bN(t)
\end{align}
Note that here we focus on isotropic fracture mechanism where the moment tensor is known save for the scaling factor. Nevertheless our approach can be extended to jointly estimate the time, location and moment tensor for other mechanisms such as double couple, \cite{Aki_Book}.

\begin{figure*}[ht]
\centering \makebox[0in]{
    \begin{tabular}{c}
      \includegraphics[width=.78\textwidth, height = 1.65in]{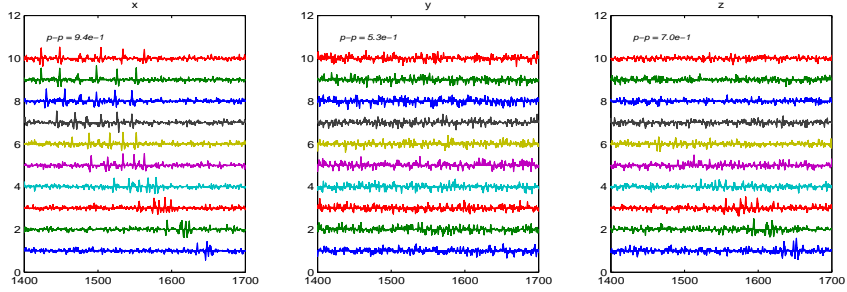}
      \end{tabular}}
  \caption{This figure shows the x,y, and z traces for all 6 synthetic events in the presence of noise.}
  \label{fig:crackPropTraces}
\end{figure*}

\vspace{-2mm}
\section{Sparsity penalized framework for HFM}
\label{sec:SPRT}
\vspace{-2mm}
To illustrate the sparsity of the HFM signal, we begin by defining a propagator, $\bGamma_{\rho}^{i} = \{\bGamma_{j\rho}^{i}(t)\}_{j = 1}^{L}$, which corresponds to noiseless data at the receivers as excited by a hypothetical seismic event $i$ (say) at location $l_i$ and time $t_i$ with an \emph{impulse} source signal, i.e.,
\begin{align}
\label{eq:propagator}
\mathbf{\Gamma}_{j\rho}^{i}(t) = \delta(t-\tau_{{\rho}_{ij}} )\ast \bP_{ij \rho}
\end{align}
With this notion of a propagator we construct a basis for a generalized Radon transform as follows. We \emph{discretize} the search volume in space and time with locations within the search volume say $V$ indexed from $1$ to $n_{V}$ and time of the events in this volume indexed from $1$ to $n_{t_V}$. Then for each of the discretized locations and time we construct a propagator using Equation~(\ref{eq:propagator}). For sake of exposition we \emph{vectorize} the space time propagators by stacking the tri-axial time traces $\bGamma_{j\rho}^{i}(t)$ for each receiver $j$  as column vectors obtaining a $3 \times T \times L$ vector for each propagator. We denote this vector by $\bGamma_{\rho}^{i}(:)$. (NOTE - here $(:)$ denotes the vectorization operation similar to the MATLAB operator $(:)$). Running over the locations and time indices over the events, we collect the vectorized propagators as columns of a matrix denoted by $\bphi_{\rho}$.   The data is also vectorized to a vector, $\bY_{\rho}(:)$ which is a $3 \times T \times L$ vector in a similar manner as the propagators.   Then a generalized discrete Radon transform of the data is given by,
\begin{align}
\label{eq:GDRT1}
R(\bY) = \bphi^{\dagger}_{\rho} \bY_{\rho}(:)
\end{align}
where
\begin{align}
\Phi_\rho=\begin{bmatrix} \bGamma_{\rho}^{1}(:), \bGamma_{\rho}^{2}(:), \hdots, \bGamma_{\rho}^{i}(:), \hdots, \bGamma_{\rho}^{n_V\times n_{t_V}}(:) \end{bmatrix}
\end{align}
and $^{\dagger}$ denotes the conjugate transpose.
This linear operator, $\bphi^{\dagger}_{\rho}$, transforms the vectorized trace data, $\bY(:)$ to a function of source location and the time at which an event occurs. It is essentially similar to generating a slant-stack (or time-moveout) map of the data, albeit along the non-linear trajectories. Note that because the noise is spatially incoherent the transform will spread the noise energy evenly across all of the possible propagators. From this transform we can see that the trace data becomes sparse when viewed in the in the generalized Radon domain (figure \ref{fig:snrTraces}).  However because the dictionary is over complete, we see a number of sidelobes for just a single seismic event, making it difficult to determine the location of the event in heavy noise. As stated in the problem statement we assume that the signal at the source is time compact. Therefore we assume that the signal $s_{i}$ for an event $i$ is supported on a set ${\cal T}_i = [t_{min}, t_{max}] \subset[0, T]$ for some total time $T$ of microseismic stimulation. In this case the source signals $s_{i_\rho}(t)$ can be viewed as weighted sum of several delta functions,
\begin{align}
\label{eq:sumDeltas}
s_{i_{\rho}}(t) = \sum_{t_k \in {\cal T}_{i_\rho}} \eta^{i}_{\rho_{t_k}} \delta(t-t_k)
\end{align}
where the vector $\eta_{t_k}$ determine the \emph{shape} of the source waveform. Here we would like to point our that in contrast to methods that assume knowledge of the signal waveform, \cite{Liu_IGARSS09}, here we have not assumed any knowledge of the wave shape.  However, we assume that the search volume contains the set ${\cal T}_{i_{\rho}}$ for the event under consideration.
\begin{align}
\label{eq:sumDeltas1}
s_{i_{\rho}}(t) =  \sum_{t_k \in [0, T]} \eta^{i}_{\rho_{t_k}} \delta(t-t_k) \,\,: \,\, \eta^{i}_{t_k} = 0 \, \forall\, t_k \notin {\cal T}_{i_{\rho}}
\end{align}
Given our definition of a propagator, assuming no geometrical spreading, we can define the arriving observed vectorized traces $\bY_\rho(:)$ due to an event as a sum of propagators with the same source location and varying event times. Therefore we can write,
\begin{align}
\label{eq:thingy6}
\bY_{\rho}(:) = \bphi_{\rho}\bX(:) + \bN
\end{align}
where $\bX(:) \in \Real^{n_V \cdot n_{t_{v}}}$ is the coefficient vector (with elements $\eta_{i}^{t_k}$ ) corresponding to the spatio-temporal volume of possible events.  If we were to reshape the true vector $\bX(:)$ into a two dimensional matrix $\bX$ the dictionary weights corresponding to a single event would be sparse along the location dimension while compact along the temporal dimension, see Figure \ref{fig:GDRT_example}.  In this way the signal exhibits sparsity across the rows of $\bX$ and is thus said to be simultaneously sparse.
\vspace{-2mm}
\section{Sparsity penalized algorithm for event localization}
\label{sec:SPRT_algo}
\vspace{-2mm}
Motivated by the row-sparse structure of the data, we use the algorithm presented in \cite{Tropp2006SPa} for a high spatio-temporal resolution mapping of the micro-seismic events. The algorithm corresponds to the following mathematical optimization problem also known as group sparse penalization, \cite{Majumdar09} in the literature.
\begin{align}
\label{eq:minl12}
\hat{\bX} = \underset{\bX}{\operatorname{\emph{argmin}}} ||\bY(:) - \bphi \bX(:)||_{2} + \lambda ||\bX||_{1,2}
\end{align}
where $\lambda$ is a sparse tuning factor that controls the group sparseness of $\bX$ versus the signal error.  Where group or row spareness is the measure of number of non-zero rows of $\bX$.  We can induce this minimizing the the $\ell_1$ norm of the euclidian norm of the rows (same time indices) of $\bX$. The parameter $\lambda$ is chosen depending on the noise level and the anticipated number of events. For the application at hand we use the method proposed in \cite{AeronTSP2011}\cite{Bose_GJI} for selection of a good regularization parameter. In this context our method is different than that used in \cite{Sacchi_GJI12} in that we are exploiting the row-sparsity of the generalized discrete radon transform.
\vspace{-1mm}
\subsection{Algorithm workflow}
\label{sec:workflow}
\vspace{-1mm}
\begin{figure*}[ht]
\centering \makebox[0in]{
    \begin{tabular}{c}
      \includegraphics[width=.98\textwidth, height = 1.6in]{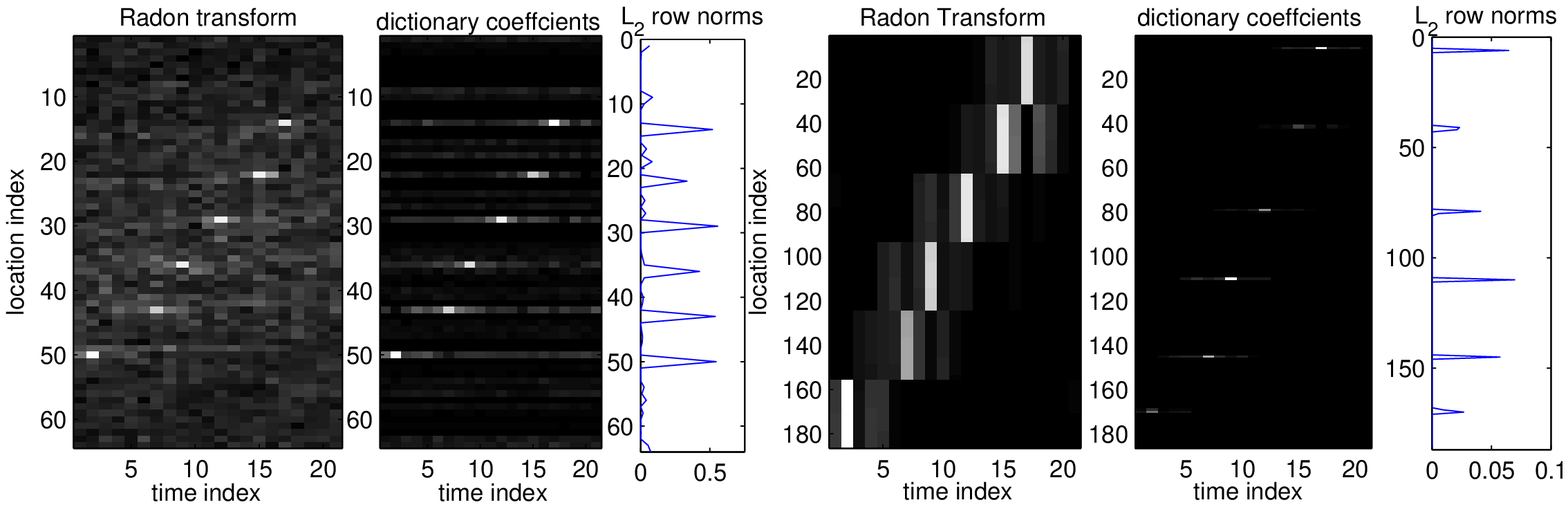}
      \end{tabular}}
  \caption{Left: This figure shows the resulting dictionary coefficients of the sparse penalized minimization for the sagittal localization and the radon transform of the observed data.   Right:  This figure shows the resulting dictionary coefficients of the sparse penalized minimization of the strike minimization and the strike radon transformed data. The events' locations are easily identifiable after applying the minimization.  The 6 distinct horizontal bands in this image correspond to the dictionary elements from the 6 different sagittal locations. }
  \label{fig:crackPropRadon}
\end{figure*}
The geometry of the linear seismic array lends itself to dividing the task of localization into two subtasks, viz.,
\vspace{0mm}
\begin{itemize}
\vspace{-1mm}
\item Localization in the sagittal plane (estimate of $\theta$ and $r$); and
\vspace{-1mm}
\item Determination of the strike angle $\phi$.
\end{itemize}
In order to determine the location of the seismic event in the sagittal plane the two horizontal axes (x and y) of the traces are combined into a single quantity $\bY_h(t)$ allowing the data to be observed in the $\bY_h(t)$ and $\bY_z(t)$ domain, $\bY'(t)$, where
\begin{align}
 \label{eq:thingy11}
 \begin{split}
 Y_{jh}(t) = \sqrt{Y_{jx}(t)^2+Y_{jy}(t)^2}
 \end{split}
 \begin{split}
 \bY_{j}'(t) =
 \begin{bmatrix}
  Y_{jh}(t)  \\
  Y_{jz}(t) \\
  \end{bmatrix}
 \end{split}
 \end{align}
Furthermore, the transformation is applied to each dictionary elements and the modified dictionary $\bphi_{hz}$ is constructed. By constructing the dictionary in this domain both the number of dictionary elements are reduced by the number of possible strike angles and decreases the size of each dictionary entry by the number of time samples.  However, because the amplitude difference across the $x$ and $y$ components are eliminated in this procedure, all events having the same strike will have the same amplitude in the radon domain.    The $\ell_{1,2}$ minimization that was described in Equation~\ref{eq:minl12}, can now be applied to the traces in the $\bY'(t)$ space. The minimization now takes the form of \ref{eq:sagEst} in which $\bphi_{hz} $ is the compressed dictionary and $\bX_{hz} $ is the dictionary weights and results in the estimation of signals time and sagittal location support, $ \hat{\bX}_{hz}$.
\begin{align}
\label{eq:sagEst}
\hat{\bX}_{hz} = \underset{\bX_{hz}}{\operatorname{\emph{argmin}}} ||\bY_{hz}(:) - \bphi_{hz} \bX_{hz}(:)||_{2} + \lambda ||\bX_{hz}||_{1,2}
\end{align}
Once the minimization has been performed the range and depth of the location can then be determined by thresholding the dictionary coefficients. To locate the events or a single event we take the $\ell_2$ norm of each row across time to generate a column vector whose entries represent the total signal energy at a given location.  The largest entries of these vectors correspond to location of the seismic event. Once a set of seismic events have been located in the sagittal plane the algorithm has identified the events' strike, range, and time with some uncertainty, the event's true location is effectively constrained to several  semi-toruses of constant range and dip as the only unknown parameter is strike.  Because of the possible locations of the event have been constrained to a small subset of the entire search volume, the computational load of performing the minimization on the full three axes traces has been significantly decreased.  In order to estimate the strike of a seismic event a subset of the original dictionary, $\bphi_{xyz}$, consisting of propagators in the full axes setup only along the semi-toruses.   Once this small dictionary is constructed a $\ell_{(1,2)}$ minimization is then applied to the full three dimensional data to estimate the signal's strike support $\hat{\bX}_{xyz}$.
 \begin{align*}
\label{eq:strike}
\hat{\bX}_{xyz} = \underset{\bX_{xyz}}{\operatorname{\emph{argmin}}} ||\bY_{xyz}(:) - \bphi_{xyz} \bX_{xyz}(:)||_{2} + \lambda ||\bX_{xyz}||_{1,2}
\end{align*}
 As with the sagittal localization, we again take the $\ell_2$ norm of each row of the dictionary coefficients ,$\hat\bX_{xyz}$ , across time to generate a column vector.  Because the number of events has been determined in sagittal localization, the strike of each event can be determined by picking the corresponding number of largest peaks of the column vector.
\vspace{-2mm}
\section{Performance Evaluation on Synthetic Data}
\label{sec:results}
\vspace{-2mm}
To illustrate the performance of the proposed method 6 seismic events were generated in short succession after with a temporal spacing of 5 milliseconds, see Figure \ref{fig:crackPropTraces}. Furthermore, the seismic events follow a near linear path occurring roughly 50 meters apart.  In this simulation a search volume with a depth from 750 to 1050 meters, a range from 750 to 1050 meters, and set of strike angles from 0 to 30 degrees.  A spatial resolution of 50 meters and .1 degrees was used.  Each of the seismic events generated a 150Hz compressional ricker wavelet with the same amplitude across all events.   Although these traces do not overlap along some of the detectors, at many detectors the events fall on top of each other, making it difficult to determine the arrival times of each event and corrupting the relatively amplitude across the x and y axes (figure \ref{fig:crackPropTraces}).  As in the multiple events simulation the same $\ell_{(1,2)}$ minimization and thresholding operations were applied to the data.  Figure \ref{fig:crackPropRadon} shows the resulting radon transform and dictionary coefficients.  After applying the minimization it easy to determine the strike and sagittal location of each of the 6 events.

\begin{figure}[h]
\centering \makebox[0in]{
    \begin{tabular}{c}
      \includegraphics[width=.48\textwidth, height = 1.5 in]{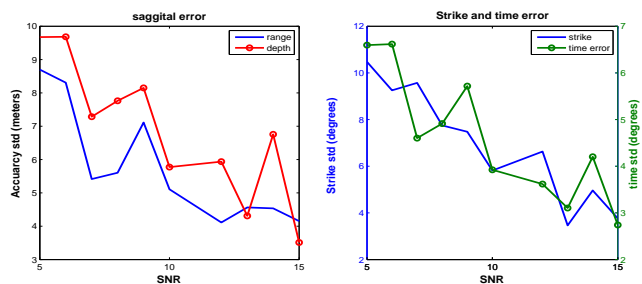}
      \end{tabular}}
  \caption{Left figure shows the accuracy in sagittal position estimation with SNR averaged over 25 synthetic traces at each SNR.  Right figure shows the accuracy in strike and time estimation of as a function of SNR. }
  \label{fig:snrAcc}
\end{figure}

In order to characterize the performance of our algorithm in the presence of noise, the simulation was repeated 25 times with a single seismic event occurring in the middle of the search volume with the same detector configuration but with varying SNR's from 1 to 15.  In addition, the sagittal resolution was increases to 5 meters.  The results of the application of the algorithm to the varying SNR cases are shown in figure \ref{fig:snrAcc}.  The algorithm is able to accurately localize seismic events in the sagittal to plane to within 10 meters for signal to noise ratios above 8.  However, below a signal to noise ratio of 5 the algorithm becomes unstable the algorithm consistently estimates the event as occurring on the edge of the search volume 50 meters away from the true event's location, driving the apparent track accuracy to 50 meters second.

\vspace{-4mm}
\section{Conclusion \& Future Work}
\label{sec:conc_future}
\vspace{-2mm}
We demonstrated a novel method based on sparsity penalized reconstruction methods for accurate HFM. Our method assumes minimal knowledge of the waveform signatures and also exploits the temporal information in the signal. Currently our processing works on the compressional and shear waves separately. In future we will extend this framework to jointly process these arrivals and exploit the dependencies in polarization to improve the accuracy.

\vspace{-2mm}
\section{Acknowledgements}
\vspace{-2mm}
The authors will like to acknowledge many helpful discussions and suggestions from Dr. Sandip Bose at Schlumberger-Doll Research, Cambridge, MA.

\vspace{-4mm}
\bibliographystyle{IEEEbib}
\bibliography{HFM_report}
\end{document}